\documentclass[12pt,a4paper]{article}

\usepackage{amssymb}

\newcommand{\mathsym}[1]{{}}

\newcommand{\ket}[1]{\left|#1\right\rangle}      

\begin{document}

\newpage
\setcounter{page}{0}
\begin{titlepage}
\begin{flushright}
{\footnotesize AEI - 2009 - 012}
\end{flushright}
\vskip 1.5cm
\begin{center}
{\Large \textbf{ The Bethe Ansatz Equations \\ for Reflecting Magnons}}\\
\vspace{1.5cm}
{\large W. Galleas} \\
\vspace{2cm}
{\it Max-Planck-Institut f\"ur Gravitationphysik \\
Albert-Einstein-Institut \\
Am M\"uhlenberg 1, 14476 Potsdam, Germany}\\
\end{center}
\vspace{1.5cm}

\begin{abstract}
We derive the Bethe Ansatz Equations on the half line for particles interacting through
factorized $S$-matrices invariant relative to the centrally extended $su(2|2)$ Lie superalgebra
and $su(1|2)$ open boundaries. These equations may be of relevance for the study of the spectrum 
of open strings on $AdS_{5} \times S^5$ background attached to $Y=0$ giant graviton branes. An one-dimensional 
spin chain hamiltonian associated to this system is also derived.
\end{abstract}

\vspace{2.0cm}
\centerline{{\small PACS numbers:  05.50+q, 02.30.IK}}
\vspace{.1cm}
\centerline{{\small Keywords: Bethe ansatz, Reflecting magnons, Open boundaries}}
\vspace{2.0cm}
\centerline{{\small February 2009}}
\end{titlepage}

\section{Introduction}

Integrability in gauge and string theories has been a subject of intensive research in the last years, where a variety of
novel integrable structures has emerged. The existence of integrable structures in gauge theories, at the classical as well as at the quantum level, has been suspected for a long time by means of different setups \cite{intro1}, but undoubtedly
it gained a whole new significance in the scenario of the AdS/CFT correspondence \cite{AdS}. In this context the celebrated Bethe ansatz acquired
a new status in contemporary physics providing the spectrum of certain string and gauge theories. 

Currently the most well studied case of gauge/string duality consist of the type IIB string theory in $AdS_{5} \times S^5$ space
together with its gauge counterpart $\mathcal{N}=4$ super Yang-Mills in four dimensions, where a large amount of evidences supports integrability. One natural
question emerging in this scenario is if the integrability of the $\mathcal{N}=4$ super Yang-Mills \cite{intro3} and of the classical
string sigma model in the $AdS_{5} \times S^5$ space \cite{intro4} is present in other theories. In the $\mathcal{N}=4$ super Yang-Mills
integrability is made manifest in the computation of anomalous dimensions of single trace operators. However, the gauge theory also
contain baryonic operators \cite{intro5} which for instance are given by determinants instead of traces. Such baryonic operators, i.e. the so called 
giant gravitons \cite{intro6}, are of particular interest in the context of AdS/CFT duality corresponding to D-brane excitations in the
string counterpart.

From the string theory point of view it is expected to have boundaries when we consider D-branes. The excitations of open strings are then
described by a two-dimensional field theory with a boundary and in the context of integrable field theories it is often possible to define the
system on a half line with suitable boundary conditions such that the system remains integrable \cite{intro7}. Such D-branes
can appear in several circumstances, for instance: conformal field theories with defects \cite{intro8}, gauge theories added with fundamental
flavours \cite{intro9} and certain baryonic operators in $\mathcal{N}=4$ super Yang-Mills \cite{intro12}. Concerning the last case, 
Berenstein and Vazquez have shown in \cite{intro12} that the one-loop mixing of non-BPS giant gravitons can be described
within the paradigm of integrable spin chains with open boundary conditions, and given the current status of higher loop integrability
for single trace operators in $\mathcal{N}=4$ super Yang-Mills \cite{intro11}, it is a reasonable goal to examine the spin
chain interpretation of the mixing of non-BPS giant gravitons beyond one-loop order. This problem was first approached in \cite{AGAR}
where the author computed the corresponding two-loop open spin chain hamiltonian. Subsequently in \cite{HM} Hofman and Maldacena
obtained a two-loop integrable open spin chain for open strings like operators coupled to maximal giant gravitons. 
Previous works studying open strings with a variety of boundary conditions and the corresponding gauge theory open spin chain
also include for instance the Refs. \cite{STEF}-\cite{DRU}.

Integrability at the boundaries is often associated with the so called boundary Yang-Baxter equation or reflection equation \cite{CHER,SK}.
In this context, the authors in \cite{HM} proposed a reflection matrix setup in order to study open strings attached to maximal
giant gravitons in $AdS_{5} \times S^5$ background. Here we shall focus on the $Y=0$ giant graviton brane case or $su(1|2)$ theory described 
in \cite{HM} and the aim of this paper is to derive the associated Bethe ansatz equations determining the energy of the
system.

This paper is organized as follows. In the next section we present the centrally extended $su(2|2)$ invariant $S$-matrix in a 
suitable basis for a further algebraic Bethe ansatz analysis. In the section 3 we describe the double-row transfer matrix with the required symmetries 
associated to the problem of open strings on $AdS_{5} \times S^5$ background attached to $Y=0$ giant graviton branes. The sections 4 and 5 are
devoted to the derivation of the associated spectrum and in the section 6 we derive a spin chain hamiltonian with open boundary conditions based on
a non-regular $su(1|2)$ reflection matrix. Concluding remarks are discussed in the section 7 and in the appendix A we establish 
the formal connection between previous results presented in the literature concerning the explicit form of the $Y=0$ reflection
matrix.

\section{The centrally extended $su(2|2)$ $S$-matrix}

Integrability in the context of planar AdS/CFT correspondence and in the one-dimensional Hubbard model are intimately connected with the
centrally extended $su(2|2)$ Lie superalgebra. As shown in \cite{tba1,tba1a,PLEF}, this algebra is strong enough to completely determine
the form of the $S$-matrix up to an overall multiplicative scalar factor. In the Ref. \cite{BK} the authors derived a $S$-matrix based on
a $q$-deformation of the centrally extended $su(2|2)$ Lie superalgebra. The above mentioned $S$-matrix satisfies the graded Yang-Baxter equation which for instance reads
\begin{equation}
\label{YB}
S_{12}(\lambda_{1},\lambda_{2}) S_{13}(\lambda_{1},\lambda_{3}) S_{23}(\lambda_{2},\lambda_{3}) = 
S_{23}(\lambda_{2}, \lambda_{3}) S_{13}(\lambda_{1} , \lambda_{3}) S_{12}(\lambda_{1} , \lambda_{2}).
\end{equation}
This equation is defined in the tensor product $V_{\lambda_{1}} \otimes V_{\lambda_{2}} \otimes V_{\lambda_{3}}$ where $V_{\lambda_{i}}$ is a finite dimensional $Z_2$ graded space 
parameterized by a rapidity $\lambda_{i}$. Each one of such spaces carries a representation $\Pi_{\lambda_{i}} : U_{q}\left[ \mathcal{G} \right] \rightarrow \mbox{End} \left( V_{\lambda_{i}} \right)$, where here $\mathcal{G}$ refers to the centrally extended $su(2|2)$ Lie superalgebra, and
$S_{ij}$ consist of a complex valued matrix $S_{ij}: \mathbb{C} \rightarrow \mbox{End}\left( V_{\lambda_{i}} \otimes V_{\lambda_{j}} \right)$ 
acting non trivially in the $i$th and $j$th spaces of $\mbox{End}\left( V_{\lambda_{1}} \otimes V_{\lambda_{2}} \otimes V_{\lambda_{3}} \right)$. 
Though our discussion will be restricted to the limit $q\rightarrow 1$, the structure of the $q$-deformed $S$-matrix presented in \cite{BK} 
has shown to be more suitable for the forthcoming analysis.

The tensor products appearing in the above definitions are understood in the graded sense.
For instance, the matrix components of $A \otimes B$ are given by
$\left[ A \otimes B  \right]^{\alpha \gamma}_{\beta \; \delta} = A^{\alpha}_{\beta} B^{\gamma}_{\delta} (-1)^{(p_{\alpha} + p_{\beta})p_{\gamma}}$
which carries an explicit dependence of the Grassmann parities $p_{\alpha}$. These Grassmann parities assume values on the group
$Z_{2}$ and enable us to characterize bosonic and fermionic degrees of freedom.
In particular, the $\alpha$th degree of freedom is distinguished by the
Grassmann parity
\begin{equation}
p_{\alpha}=\cases{
0  \;\;\;\;\; \mbox{for} \;\; \alpha \;\; \mbox{bosonic} \cr
1  \;\;\;\;\; \mbox{for} \;\; \alpha \;\; \mbox{fermionic} \cr }.
\end{equation}

We shall consider the limit $q\rightarrow 1$ of the $S$-matrix given in \cite{BK} adopting the grading structure $BFFB$.
More precisely, the Grassmann parities are set to
\begin{equation}
\label{grad}
p_{\alpha}=\cases{
0  \;\;\;\;\; \alpha = 1,4 \cr
1  \;\;\;\;\; \alpha = 2,3 \cr }
\end{equation}
and the centrally extended $su(2|2)$ invariant $S$-matrix can be written as
{\scriptsize
\begin{eqnarray}
\label{smat}
S(\lambda_{1}, \lambda_{2}) = 
\pmatrix{
a_{1}  & 0  & 0 & 0 & 0  & 0 & 0 & 0 & 0 & 0 & 0 & 0 & 0 & 0 & 0 & 0\cr
0 & a_{8} & 0 & 0 & a_{10} & 0 & 0 & 0 & 0 & 0 & 0 & 0 & 0 & 0 & 0 & 0 \cr
0 & 0 & a_{8} & 0 & 0 & 0 & 0 & 0 &a_{10} & 0 & 0 & 0 & 0 & 0 & 0 & 0 \cr
0 & 0 & 0 & a_{3}  & 0 & 0 & a_{7}  & 0 & 0 & -a_{7}  & 0 & 0 & a_{2}  & 0 & 0 & 0 \cr
0 & a_{9}  & 0 & 0 & a_{11}  & 0 & 0 & 0 & 0 & 0 & 0 & 0 & 0 & 0 & 0 & 0 \cr 
0 & 0 & 0 & 0 & 0 & a_{4}  & 0 & 0 & 0 & 0 & 0 & 0 & 0 & 0 & 0 & 0 \cr
0 & 0 & 0 & -a_{7}  & 0 & 0 & a_{6}  & 0 & 0 & a_{5}  & 0 & 0 & a_{7}  & 0 & 0 & 0 \cr
0 & 0 & 0 & 0 & 0 & 0 & 0 & a_{11}  & 0 & 0 & 0 & 0 & 0 & a_{9}  & 0 & 0 \cr
0 & 0 & a_{9}  & 0 & 0 & 0 & 0 & 0 & a_{11}  & 0 & 0 & 0 & 0 & 0 & 0 & 0 \cr
0 & 0 & 0 & a_{7}  & 0 & 0 & a_{5}  & 0 & 0 & a_{6}  & 0 & 0 & -a_{7}  & 0 & 0 & 0 \cr
0 & 0 & 0 & 0 & 0 & 0 & 0 & 0 & 0 & 0 & a_{4} & 0 & 0 & 0 & 0 & 0  \cr
0 & 0 & 0 & 0 & 0 & 0 & 0 & 0 & 0 & 0 & 0 & a_{11}  & 0 & 0 & a_{9}  & 0 \cr
0 & 0 & 0 & a_{2}  & 0 & 0 & -a_{7}  & 0 & 0 & a_{7}  & 0 & 0 & a_{3}  & 0 & 0 & 0 \cr
0 & 0 & 0 & 0 & 0 & 0 & 0 & a_{10}  & 0 & 0 & 0 & 0 & 0 & a_{8}  & 0 & 0 \cr
0 & 0 & 0 & 0 & 0 & 0 & 0 & 0 & 0 & 0 & 0 & a_{10}  & 0 & 0 & a_{8}  & 0 \cr
0 & 0 & 0 & 0 & 0 & 0 & 0 & 0 & 0 & 0 & 0 & 0 & 0 & 0 & 0 & a_{1}  \cr}.
\end{eqnarray}}
The $36$ non-null entries are parameterized by complex variables $x^{\pm}_{j} = x^{\pm}(\lambda_{j})$ constrained by the relations
\begin{equation}
\label{eqv}
\frac{x^{+}_{j}}{x^{-}_{j}} = e^{i \lambda_{j}} \;\;\; \mbox{and} \;\;\; 
x^{+}_{j} + \frac{1}{x^{+}_{j}} - x^{-}_{j} - \frac{1}{x^{-}_{j}} = \frac{i}{g} ,
\end{equation}
where in the context of the AdS/CFT duality, $g$ corresponds to the string sigma model coupling constant and $\lambda_{j}$ denote the world-sheet rapidities.
With the above considerations, the entries of (\ref{smat}) can be explicitly written in terms
of the variables $x^{\pm}_{j}$ as
\begin{eqnarray}
\label{bw}
a_{1} &=& \left(\frac{x^{+}_{1} x^{-}_{2}}{x^{-}_{1} x^{+}_{2}} \right)^{\frac{1}{2}} \frac{(x^{+}_{2} - x^{-}_{1})}{(x^{-}_{2} - x^{+}_{1})}
\;\;\;\;\;\;\;\;\;\;\;\;\;\;\;\;\;\;\;\;\;\;\;\;\;\;\;\;\;\;
a_{2} = \left(\frac{x^{+}_{1} x^{-}_{2}}{x^{-}_{1} x^{+}_{2}} \right)^{\frac{1}{2}} \frac{(x^{+}_{1} - x^{-}_{1})}{(x^{-}_{2} - x^{+}_{1})}
\frac{(x^{+}_{1} x^{-}_{1} x^{-}_{2} - x^{+}_{2})}{(x^{-}_{1} x^{-}_{2} -1) x^{+}_{1}}   \nonumber \\
a_{3} &=& \frac{x^{-}_{1}}{x^{+}_{1}} \left( \frac{x^{+}_{1} x^{-}_{2}}{x^{-}_{1} x^{+}_{2}} \right)^{\frac{1}{2}} \frac{(x^{+}_{1} - x^{+}_{2})}{(x^{+}_{1} - x^{-}_{2})}
\frac{(x^{-}_{2} x^{+}_{1} - 1)}{(x^{-}_{1} x^{-}_{2} -1)}
\;\;\;\;\;\;\;
a_{4} = 1 \nonumber \\
a_{5} &=& \frac{(x^{+}_{2} - x^{-}_{2})}{(x^{+}_{1} - x^{-}_{2})} \frac{(x^{+}_{2} x^{-}_{1} x^{-}_{2} - x^{+}_{1})}{(x^{-}_{1} x^{-}_{2} - 1) x^{+}_{2}}
\;\;\;\;\;\;\;\;\;\;\;\;\;\;\;\;\;\;\;\;\;\;\;
a_{6} = \frac{x^{-}_{2}}{x^{+}_{2}} \frac{(x^{+}_{1} - x^{+}_{2})}{(x^{+}_{1} - x^{+}_{2})} \frac{(x^{-}_{1} x^{+}_{2} - 1)}{(x^{-}_{1} x^{-}_{2} - 1)}
\nonumber \\
a_{7} &=& \frac{x^{-}_{1} x^{-}_{2}}{x^{+}_{1} x^{+}_{2}} \left( \frac{x^{+}_{1}}{x^{-}_{1}} \right)^{\frac{1}{2}} \frac{(x^{+}_{1} - x^{+}_{2} ) \gamma_{1} \gamma_{2}}{(x^{-}_{2} - x^{+}_{1})(x^{-}_{1} x^{-}_{2} - 1)} \;\;\;\;\;\;\;\;
a_{8} = \left( \frac{x^{+}_{2}}{x^{-}_{2}} \right)^{-\frac{1}{2}} \frac{(x^{+}_{2} - x^{+}_{1})}{(x^{-}_{2} - x^{+}_{1})}
\nonumber \\
a_{9} &=& \frac{(x^{+}_{2} - x^{-}_{2})}{(x^{-}_{2} - x^{+}_{1})} \frac{\gamma_{1}}{\gamma_{2}} \;\;\;\;\;\;\;\;\;\;\;\;\;\;\;\;\;\;\;\;\;\;\;\;\;\;\;\;\;\;\;\;\;\;\;\;\;\;\;\;\;\;\;
a_{10} = \left( \frac{x^{+}_{1} x^{-}_{2}}{x^{-}_{1} x^{+}_{2}} \right)^{\frac{1}{2}} \frac{(x^{+}_{1} - x^{-}_{1})}{(x^{-}_{2} - x^{+}_{1})} \frac{\gamma_{2}}{\gamma_{1}}
\nonumber \\
a_{11} &=& \left( \frac{x^{+}_{1}}{x^{-}_{1}} \right)^{\frac{1}{2}} \frac{(x^{-}_{2} - x^{-}_{1})}{(x^{-}_{2} - x^{+}_{1})}
\end{eqnarray}
where $\gamma_{j} = \sqrt{-i \left(\frac{x^{+}_{j}}{x^{-}_{j}} \right)^{\frac{1}{2}} (x^{+}_{j} - x^{-}_{j})}$.
In addition to the Yang-Baxter relation (\ref{YB}), the $S$-matrix defined by (\ref{smat}-\ref{bw}) also fulfills the following properties
\begin{eqnarray}
\label{prop}
\mbox{{\bf Regularity}} &:&  S_{12}(\lambda,\lambda) = - P_{12} \nonumber \\
\mbox{{\bf Unitarity}} &:&  S_{12}(\lambda_{1} , \lambda_{2}) S_{21}(\lambda_{2},\lambda_{1}) = I_{12}
\end{eqnarray}
where $P$ and $I$ denote respectively the graded permutation operator and the identity matrix. As we shall see in the next sections these properties will be of
relevance for the construction of integrable systems with open boundaries.

\section{Double-row transfer matrix and reflection matrices}

The generalization of the Quantum Inverse Scattering Method for systems with open boundaries
proposed by Sklyanin \cite{SK} gave a large impulse to the study of integrable systems with non-periodic boundary conditions. 
In Sklyanin's formalism the construction of integrable models with open boundaries is based on the solutions of 
the so called reflection equations for a given integrable bulk system.

In order to consider the centrally extended $su(2|2)$ model some generalizations
of Sklyanin's original approach have to be taken into account due to some particular features of the $S$-matrix.
Firstly the $S$-matrix (\ref{smat}-\ref{bw}) does not depend on the difference 
of the spectral parameters and in fact, it is equivalent to Shastry's $R$-matrix \cite{tba1a,MELO} embedding the one-dimensional
Hubbard model \cite{Sha}. This feature leads us to consider the approach proposed in \cite{ZHOU,BRACK} and subsequently considered by other authors \cite{WAD}.
It turns out that an inhomogeneous transfer matrix with open boundaries can be written as the following supertrace over the
auxiliar space $\mathcal{A} \equiv \mathbb{C}^4$,
\begin{equation}
\label{tm}
T(\lambda , \{\lambda_{j} \}) = \mbox{Str}_{\mathcal{A}} \left[ K^{+}_{\mathcal{A}}(\lambda) \mathcal{T}_{\mathcal{A}}(\lambda, \{ \lambda_{j} \} )
K^{-}_{\mathcal{A}}(\lambda) \mathcal{T}^{-1}_{\mathcal{A}}(-\lambda, \{ \lambda_{j} \} ) \right]
\end{equation}
where $\lambda$ is the world-sheet rapidity which parameterizes the integrable manifold \footnote{We recall here that the variable $\lambda$ is also related to variables
$x^{\pm}$ through the relations (\ref{eqv}).} and the set of variables
$\lambda_{1} , \dots , \lambda_{N}$ denotes the inhomogeneities. The matrix 
\begin{equation}
\mathcal{T}_{\mathcal{A}}(\lambda, \{ \lambda_{j} \} )= S_{\mathcal{A} N}(\lambda, \lambda_{N}) \dots S_{\mathcal{A} 1}(\lambda, \lambda_{1})
\end{equation}
is the standard monodromy matrix which generates the corresponding closed chain with $N$ sites while
$\mathcal{T}^{-1}_{\mathcal{A}}(-\lambda, \{ \lambda_{j} \} )$ is given by
\begin{equation}
\label{imono}
\mathcal{T}^{-1}_{\mathcal{A}}(-\lambda, \{ \lambda_{j} \} ) = S_{1 \mathcal{A} }(\lambda_{1}, -\lambda) \dots S_{N \mathcal{A} }(\lambda_{N}, -\lambda)
\end{equation}
due to the unitarity property (\ref{prop}). In their turn the matrices $K^{\pm}_{\mathcal{A}}(\lambda)$ describe the interactions at the right and left
ends of the open chain. 

We remark here that an equivalent transfer matrix was constructed previously in the literature for open strings attached to maximal giant
gravitons \cite{NEPO}. However, we shall consider here a different basis which results in a transfer matrix more suitable for an 
algebraic Bethe ansatz analysis.

Integrability at the boundaries is governed by the so called reflection equations. Within the graded version of the Quantum Inverse Scattering
Method the matrix $K^{-}(\lambda)$ is required to satisfy
\begin{eqnarray}
\label{rem}
S_{12}(\lambda,\mu) K_{1}^{-}(\lambda) S_{21}(\mu,-\lambda) K_{2}^{-}(\mu) = 
K_{2}^{-}(\mu) S_{12}(\lambda,-\mu) K_{1}^{-}(\lambda) S_{21}(-\mu,-\lambda)
\end{eqnarray}
while the reflection at the opposite boundary is subjected to the dual relation
\begin{eqnarray}
\label{rep}
&& S_{21}^{st_{1} ist_{2}}(\mu,\lambda) K_{1}^{+}(\lambda)^{st_{1}} S_{12}^{st_{1} ist_{2}}(-\lambda,\mu) K_{2}^{+}(\mu)^{ist_{2}}  \nonumber \\
&& = K_{2}^{+}(\mu)^{ist_{2}} S_{21}^{st_{1} ist_{2}}(-\mu,\lambda) K_{1}^{+}(\lambda)^{st_{1}} S_{12}^{st_{1} ist_{2}}(-\lambda,-\mu) ,
\end{eqnarray}
where the symbols $st_{\alpha}$ and $ist_{\alpha}$ stand respectively for the operations of supertranposition
in the space with index $\alpha$ and its inverse operation as described in \cite{BRACK}.
The role played by the reflection equations for the boundaries is similar to the one played by the Yang-Baxter equation for 
the bulk of the system and when the reflection matrices $K^{\pm}_{\mathcal{A}}(\lambda)$
satisfy (\ref{rem}) and (\ref{rep}), the double-row operator (\ref{tm}) constitutes an one parameter family of commutative transfer matrices, i.e. 
\begin{equation}
\left[ T(\lambda , \{\lambda_{j} \}) , T(\mu , \{\lambda_{j} \})  \right]=0  \;\;\;\;\; \forall \lambda , \mu \in \mathbb{C} .
\end{equation}
The commutativity of the transfer matrices for all values of the spectral parameters $\lambda$ and $\mu$
provides a complete set of operators in involution and thus ensures the integrability of the system. 

In the past much work was devoted to develop a systematic quantum group approach enabling us to find solutions of the reflection equations.
The studies on boundary quantum groups were initiated in \cite{BQG} and have been carried out since then in order to unveil the fundamental structure of their 
generators \cite{BQG1}. In \cite{BAS} it was shown that the boundary quantum group structure behind the reflection equation associated to the 
$U_{q} \left[ sl(2)  \right]$ $S$-matrix is actually a $q$-deformed Dolan-Grady algebra invariant under the coproduct homomorphism of $U_{q} \left[ sl(2)  \right]$.
However, for higher rank affine Lie algebras the analogue of such algebraic relations remains an open question. 
From the physical picture it is expected the presence of boundaries to break the quantum group symmetry of the bulk down to a certain
subgroup. This unbroken residual symmetry should be powerful enough to determine the reflection matrix as the original symmetry does for the bulk $S$-matrix.
In the Ref. \cite{DELMAC} the authors proposed a general framework for obtaining solutions of the reflection equation by solving a intertwining relation of certain 
coideal subalgebras of the symmetry algebra intertwined by the $S$-matrix. 
Let then $\mathcal{G}$ be a Hopf algebra with generators denoted by $\mathcal{Q}$ whose coproduct homomorphism $\Delta : \mathcal{G} \rightarrow \mathcal{G} \otimes \mathcal{G}$ is intertwined by the 
$S$-matrix
\begin{equation}
\label{rtt}
S \Delta(\mathcal{Q}) = \Delta_{op}(\mathcal{Q}) S \;\;\;\;\;\;\; \forall \; \mathcal{Q} \in \mathcal{G}
\end{equation}
where $\Delta_{op}$ stands for the opposite coproduct $\Delta_{op}(\mathcal{Q}) = P \Delta(\mathcal{Q}) P$. In addition to that let us also consider the
following intertwining relation
\begin{equation}
\label{ktt}
K \mathcal{Q} = \bar{\mathcal{Q}} K \;\;\;\;\;\;\;\;\;\;\;\;\;\;\;\;\;\;\; \forall \; \mathcal{Q} \in \mathcal{B} \; , \; \mathcal{B} \subset \mathcal{G}
\end{equation}
where $\bar{\mathcal{Q}}$ corresponds to the reflected generator $\mathcal{Q}$, i.e. $\lambda \rightarrow -\lambda$, and 
$\mathcal{B}$ is a left coideal subalgebra of $\mathcal{G}$. The algebra $\mathcal{B}$ is also refereed as quantum affine reflection algebra and it was shown
in \cite{DELMAC} that the $K$-matrices solving the intertwining relation (\ref{ktt}) render solutions of the reflection equation (\ref{rem}).
 
In order to consider the case of open strings on $AdS_{5} \times S^{5}$ background attached to $Y=0$ branes, we
regard $\mathcal{G}$ as the centrally extended $su(2|2)$ Lie superalgebra in its fundamental four-dimensional representation and the left coideal subalgebra
$\mathcal{B}$ as the $su(1|2)$ superalgebra, as described in \cite{HM}. The corresponding $S$-matrix is given by the relations (\ref{smat}-\ref{bw})
and in order to study the associated reflection matrices it is necessary to first determine how the variables $x^{\pm}$ behave under reflection. 

In \cite{HM} the authors showed that the requirement of energy conservation when $\lambda \rightarrow - \lambda$, preserving the constraint (\ref{eqv}),
restrict us to the mapping $x^{\pm} \rightarrow -x^{\mp}$.
This mapping together with the representation of the states involved given in \cite{BK} allow us to
use the intertwining relation (\ref{ktt}) in order to determine the reflection matrix $K^{-}(\lambda)$ up to an overall phase factor which we shall omit at
the moment. It turns out that the $su(1|2)$ reflection matrix is then given by
\begin{equation}
\label{km}
K^{-}(\lambda) = \pmatrix{
- e^{-i \frac{\lambda}{2}} & 0 & 0 & 0 \cr
0 & 1 & 0 & 0 \cr
0 & 0 & 1 & 0 \cr
0 & 0 & 0 & e^{i\frac{\lambda}{2}} \cr}.
\end{equation}
Strictly speaking, the approach devised in \cite{DELMAC} considers only non-graded algebras. However, the extension of the proposed method
to $Z_{2}$ graded algebras is expected to follow the same lines of \cite{BR} for the solutions of the graded Yang-Baxter equation based on superalgebras.
On the other hand one could have considered the direct resolution of the reflection equation (\ref{rem}) similarly to the analysis performed by Shiroishi and
Wadati in \cite{WAD} for the Hubbard model. By doing so one finds that the class of diagonal solution (\ref{km}) does
not accommodate free parameters. The direct inspection of the dual reflection equation (\ref{rep}) results in
\begin{equation}
\label{kp}
K^{+}(\lambda) =  K^{-}(-\lambda)
\end{equation}
in agreement with the reflection symmetry of the problem discussed in \cite{HM} and the crossing like relation automorphism employed in \cite{NEPO}.

Though they are equivalent, the $K$-matrix (\ref{km}) is slightly different from the one derived in \cite{HM} and the one obtained in \cite{NEPOa} by
means of the boundary Faddeev-Zamolodchikov algebra. The differences amount to a different choice of the grading structure and the gauge in which the $S$-matrix is considered. 
In the Appendix A their relationship are made precise.

\section{The Bethe ansatz approach}
The purpose of this section is to determine the spectrum of the double-row transfer matrix defined by the Eqs. (\ref{smat}-\ref{bw}),
(\ref{tm}-\ref{imono}), (\ref{km}) and  (\ref{kp}). 
In this case the boundary elements are diagonal and such eigenvalue problem can be tackled by the algebraic Bethe ansatz in the same lines
employed in the Refs. \cite{Guan,Guang}. 

The algebraic Bethe ansatz method was initially conceived for systems with periodic boundary conditions
but later on systems with more general boundary conditions were also included in this framework as well. The basic ingredients are 
still the existence of a pseudovacuum state and appropriate commutation rules between the elements of the associated monodromy matrix. 

Due to the Yang-Baxter relation (\ref{YB}) and the reflection equations (\ref{rem}), one can demonstrate within Sklyanin's approach
that the double-row monodromy matrix
\begin{equation}
\label{mono}
U_{\mathcal{A}} (\lambda , \{ \lambda_{j} \} ) = \mathcal{T}_{\mathcal{A}}(\lambda, \{ \lambda_{j} \} ) K^{-}_{\mathcal{A}}(\lambda) \mathcal{T}^{-1}_{\mathcal{A}}(-\lambda, \{ \lambda_{j} \} )
\end{equation}
satisfy the following quadratic algebra
\begin{eqnarray}
\label{arem}
S_{12}(\lambda,\mu) U_{1}(\lambda , \{ \lambda_{j} \}) S_{21}(\mu,-\lambda) U_{2}(\mu , \{ \lambda_{j} \}) = 
U_{2}(\mu , \{ \lambda_{j} \}) S_{12}(\lambda,-\mu) U_{1}(\lambda , \{ \lambda_{j} \}) S_{21}(-\mu,-\lambda), \nonumber \\
\end{eqnarray}
where here $U_{\mathcal{A}} (\lambda , \{ \lambda_{j} \} )$ consist of $4 \times 4$ matrix on the auxiliary space $\mathcal{A} \equiv \mathbb{C}^{4}$ with elements acting on the tensor product $\displaystyle \bigotimes_{j=1}^{N} \mathbb{C}^{4}$.

As a first step to establish an algebraic Bethe ansatz analysis we remark that the diagonal structure of the reflection matrices (\ref{km},\ref{kp}) permit us to use
the standard ferromagnetic state 
\begin{equation}
\label{ref}
\ket{\Psi_{0}} = \bigotimes_{j=1}^{N} \ket{0} \;\;\;\; \mbox{where} \;\;\;\; 
\ket{0}=\pmatrix{
1 \cr
0 \cr
0 \cr
0 \cr}
\end{equation}
as pseudovacuum state. The existence of a pseudovacuum state does not guarantee that we can successfully apply the algebraic
Bethe ansatz and as another requirement we still need to be able to disentangle the relation (\ref{arem}) into appropriate commutation
rules for the elements of the monodromy matrix.

For instance, we need to find among the elements of $U_{\mathcal{A}} (\lambda , \{ \lambda_{j} \} )$
the operators playing the role of creation and annihilation fields with respect to the state $\ket{\Psi_{0}}$. The previous works 
on the algebraic Bethe ansatz for the centrally extended $su(2|2)$ algebra \cite{MELO} and for the Hubbard model with open
boundaries \cite{Guan} suggest us to represent
\begin{equation}
U_{\mathcal{A}} (\lambda , \{ \lambda_{j} \} ) = \pmatrix{
B(\lambda , \{ \lambda_{j} \} ) & B_{1}(\lambda , \{ \lambda_{j} \} ) & B_{2} (\lambda , \{ \lambda_{j} \} ) & F(\lambda , \{ \lambda_{j} \} ) \cr
C_{1}(\lambda , \{ \lambda_{j} \} ) & A_{11}(\lambda , \{ \lambda_{j} \} ) & A_{12}(\lambda , \{ \lambda_{j} \} ) & B_{1}^{*} (\lambda , \{ \lambda_{j} \} ) \cr
C_{2}(\lambda , \{ \lambda_{j} \} ) & A_{21}(\lambda , \{ \lambda_{j} \} ) & A_{22}(\lambda , \{ \lambda_{j} \} ) & B_{2}^{*} (\lambda , \{ \lambda_{j} \} ) \cr
F^{*}(\lambda , \{ \lambda_{j} \} ) & C_{1}^{*}(\lambda , \{ \lambda_{j} \} ) & C_{2}^{*}(\lambda , \{ \lambda_{j} \} ) & D(\lambda , \{ \lambda_{j} \} ) \cr} 
\end{equation}
in order to depict appropriate creation and annihilation fields.

Now we can turn our attention to the eigenvalue problem,
\begin{equation}
\label{eig}
T( \lambda , \{ \lambda_{j} \} ) \ket{\Psi} = \Lambda ( \lambda , \{ \lambda_{j} \} ) \ket{\Psi} ,
\end{equation}
for the double-row transfer matrix. In the framework of the boundary algebraic Bethe ansatz, this eigenvalue problem is more conveniently written in terms
of shifted operators
\begin{eqnarray}
\label{shif}
\tilde{A}_{\alpha \beta} ( \lambda , \{ \lambda_{j} \} ) &=& A_{\alpha \beta} ( \lambda , \{ \lambda_{j} \} ) - \delta_{\alpha \beta} f_{1}(\lambda) B( \lambda , \{ \lambda_{j} \} ) \nonumber \\
\tilde{D} ( \lambda , \{ \lambda_{j} \} ) &=& D( \lambda , \{ \lambda_{j} \} ) - f_{2}(\lambda) B( \lambda , \{ \lambda_{j} \} )
+ f_{1}(\lambda) \sum_{\alpha=1}^{2} A_{\alpha \alpha} ( \lambda , \{ \lambda_{j} \} ) 
\end{eqnarray}
where $f_{1}(\lambda) = \frac{x^{-} - x^{+}}{2\left( x^{+} x^{-} \right)^{\frac{1}{2}}}$ and $f_{2}(\lambda) = \frac{( x^{-} - x^{+}) [ (x^{-})^{2} - 1 ] }{2 x^{-} (x^{+} x^{-} + 1)}$.
Taking into account the Grassmann parities (\ref{grad}) and the shifted operators (\ref{shif}), the Eq. (\ref{eig}) reads
\footnote{We recall here that the supertrace of a $n\times n$ matrix $A$ is given by $\mbox{Str}(A) = \sum_{\alpha=1}^{n} (-1)^{p_{\alpha}} A_{\alpha \alpha}$.}
\begin{equation}
\label{eig1}
 \left[ \omega_{1}^{+}(\lambda) B(\lambda , \{ \lambda_{j} \} ) + \omega_{2}^{+}(\lambda) \sum_{i=1}^{2} \tilde{A}_{\alpha \alpha}(\lambda , \{ \lambda_{j} \} ) 
+ \omega_{3}^{+}(\lambda) \tilde{D}(\lambda , \{ \lambda_{j} \} ) \right] \ket{\Psi} = \Lambda ( \lambda , \{ \lambda_{j} \} ) \ket{\Psi} , \nonumber \\
\end{equation}
where the functions $\omega_{\alpha}^{+}(\lambda)$ are given by
\begin{eqnarray}
\omega_{1}^{+} (\lambda) &=& - \frac{(x^{-})^{\frac{1}{2}} (x^{+} + x^{-}) [ 1 + (x^{+})^{2} ]}{2 (x^{+})^{\frac{3}{2}} (1 + x^{+} x^{-} ) } \nonumber \\
\omega_{2}^{+} (\lambda) &=& - \frac{x^{+} + x^{-}}{2 x^{+}} \;\; , \;\; 
\omega_{3}^{+} (\lambda) = ( \frac{x^{-}}{x^{+}} )^{\frac{1}{2}}. 
\end{eqnarray}

As we can see from (\ref{eig1}) the diagonal elements of the monodromy matrix constitutes the transfer matrix eigenvalue problem and
within the framework of the algebraic Bethe ansatz we expect the off-diagonal elements to play the role of creation and annihilation fields. 
With respect to the pseudovacuum state $\ket{\Psi_{0}}$, the diagonal elements of $U_{\mathcal{A}} (\lambda , \{ \lambda_{j} \} )$ satisfy the 
following relations
\begin{eqnarray}
\label{vac}
B(\lambda , \{ \lambda_{j} \} ) \ket{\Psi_{0}} &=&  \omega_{1}^{-}(\lambda) \prod_{j=1}^{N} a_{1}(\lambda , \lambda_{j}) a_{1}(\lambda_{j}, -\lambda)  \ket{\Psi_{0}} \nonumber \\
A_{\alpha \alpha}(\lambda , \{ \lambda_{j} \} ) \ket{\Psi_{0}} &=& \omega_{2}^{-}(\lambda) \prod_{j=1}^{N} a_{11}(\lambda , \lambda_{j}) a_{8}(\lambda_{j}, -\lambda)  \ket{\Psi_{0}} \\
D(\lambda , \{ \lambda_{j} \} ) \ket{\Psi_{0}} &=&  \omega_{3}^{-}(\lambda) \prod_{j=1}^{N} a_{3}(\lambda , \lambda_{j}) a_{3}(\lambda_{j}, -\lambda)  \ket{\Psi_{0}}  \nonumber
\end{eqnarray}
with
\begin{eqnarray}
\omega_{1}^{-} (\lambda) &=& - ( \frac{x^{-}}{x^{+}} )^{\frac{1}{2}} \;\;\; , \;\;\; 
\omega_{2}^{-} (\lambda) = \frac{x^{+} + x^{-}}{2 x^{+}} \nonumber \\
\omega_{3}^{-} (\lambda) &=& \frac{(x^{+} + x^{-})[1+(x^{-})^2]}{2(x^{+} x^{-})^{\frac{1}{2}} (1 + x^{+} x^{-})},
\end{eqnarray}
while some of the off-diagonal elements exhibit the annihilation properties
\begin{eqnarray}
\label{vac1}
C_{\alpha}(\lambda , \{ \lambda_{j} \} ) \ket{\Psi_{0}} &=& C_{\alpha}^{*}(\lambda , \{ \lambda_{j} \} ) \ket{\Psi_{0}} = 0  \nonumber \\
A_{12}(\lambda , \{ \lambda_{j} \} ) \ket{\Psi_{0}} &=& A_{21}(\lambda , \{ \lambda_{j} \} ) \ket{\Psi_{0}} = 0  \\
F^{*}(\lambda , \{ \lambda_{j} \} ) \ket{\Psi_{0}} &=& 0 \nonumber
\end{eqnarray}
in virtue of the definitions (\ref{smat}), (\ref{km}), (\ref{mono}) and (\ref{ref}).

Concerning the remaining operators $B_{\alpha}(\lambda , \{ \lambda_{j} \} )$,
$B_{\alpha}^{*} (\lambda , \{ \lambda_{j} \} )$ and $F(\lambda , \{ \lambda_{j} \} )$, they shall be regarded as creation fields with respect to the
state $\ket{\Psi_{0}}$. Moreover, the properties (\ref{vac}) and (\ref{vac1}) imply that the pseudovacuum state $\ket{\Psi_{0}}$ is one of the
eigenvectors of the double-row transfer matrix. 
Following the standard procedure of the algebraic Bethe ansatz the next task is to look for
other transfer matrix eigenvectors as linear combinations of products of creation fields acting on the pseudovacuum state $\ket{\Psi_{0}}$.
Such construction depends dramatically on the structure of the $S$-matrix considered and since the main structure of (\ref{smat}) resembles 
that of the Hubbard model, one can expect that this construction will be similar to that presented in \cite{Guan}. Considering that there are
no significant changes from the construction devised in \cite{Guan} we shall restrict ourselves to present only the final expression
for the transfer matrix eigenvalues. In terms of the variables $x^{\pm}$, it turns out that the eigenvalues $\Lambda(\lambda , \{ \lambda_{j} \} )$ are given by
\begin{eqnarray}
\label{eigl}
&& \Lambda(\lambda , \{ \lambda_{j} \} ) = \frac{x^{-} (x^{+} + x^{-}) [ 1 + (x^{+})^2 ] }{2 (x^{+})^2 (1+x^{+} x^{-})} 
\prod_{j=1}^{N} \frac{x^{+}}{x^{-}} \frac{(x^{-} + x^{-}_{j})(x^{-} - x^{+}_{j})}{(x^{+} - x^{-}_{j})(x^{+} + x^{+}_{j})}
\prod_{k=1}^{m_{0}} \frac{x^{-}}{x^{+}} \frac{[(x^{+})^2 - (z^{-}_{k})^{2}]}{[(x^{-})^2 - (z^{-}_{k})^{2}]} \nonumber \\
&&- \left[  \frac{x^{+} + x^{-}}{2 x^{+}}  \right]^2 \left[  \frac{x^{-} + \frac{1}{x^{-}} + \frac{i}{g} }{x^{-} + \frac{1}{x^{-}} + \frac{i}{2g}}  \right] 
\prod_{j=1}^{N} \frac{x^{+}}{x^{-}} \frac{(x^{-} + x^{+}_{j})(x^{-} - x^{-}_{j})}{(x^{+} + x^{+}_{j})(x^{+} - x^{-}_{j})}
\prod_{k=1}^{m_{0}} \frac{x^{-}}{x^{+}} \left[ \frac{ (x^{+})^2 - (z^{-}_{k})^2 }{(x^{-})^2 - (z^{-}_{k})^2}   \right] \nonumber \\
&& \;\;\;\;\;\;\;\;\;\;\;\;\;\;\;\;\;\;\;\;\;\;\;\;\;\;\;\;\;\;\;\;\;\;\;\;\;\;\;\;\;\;\;\;\;\;\;\;\;\;\;\;\;\;\;\;  
\times \prod_{l=1}^{n_{0}} \frac{ (\tilde{\lambda}_{l} + x^{-} + \frac{1}{x^{-}} - \frac{i}{2g} ) }{ (\tilde{\lambda}_{l} - x^{-} - \frac{1}{x^{-}} - \frac{i}{2g} ) }
\frac{( \tilde{\lambda}_{l} - x^{-} - \frac{1}{x^{-}} + \frac{i}{2g} )}{( \tilde{\lambda}_{l} + x^{-} + \frac{1}{x^{-}} + \frac{i}{2g}) }
\nonumber \\
&&- \left[  \frac{x^{+} + x^{-}}{2 x^{+}}  \right]^2 \left[  \frac{x^{-} + \frac{1}{x^{-}} }{x^{-} + \frac{1}{x^{-}} + \frac{i}{2g}}  \right] 
\prod_{j=1}^{N}  \frac{x^{+}}{x^{-}} \frac{(x^{-} + x^{+}_{j})(x^{-} - x^{-}_{j})}{(x^{+} + x^{+}_{j})(x^{+} - x^{-}_{j})} \nonumber \\
&& \;\;\;\;\;\;\;\; \times \prod_{k=1}^{m_{0}} \frac{x^{-}}{x^{+}} \left[ \frac{ (x^{+})^2 - (z^{-}_{k})^2 }{(x^{-})^2 - (z^{-}_{k})^2}   \right]
\frac{( x^{-} + \frac{1}{x^{-}} + z^{-}_{k} + \frac{1}{z^{-}_{k}} )}{( x^{-} + \frac{1}{x^{-}} + z^{-}_{k} + \frac{1}{z^{-}_{k}} + \frac{i}{g} )}
\frac{( x^{-} + \frac{1}{x^{-}} - z^{-}_{k} - \frac{1}{z^{-}_{k}} )}{( x^{-} + \frac{1}{x^{-}} - z^{-}_{k} - \frac{1}{z^{-}_{k}} + \frac{i}{g} )} \nonumber \\
&& \;\;\;\;\;\;\;\;\;\;\;\;\;\;\;\;\;\;\;\;\;\;\;\;\;\;\;\;\;\;\;\;\;\;\;\;\;\;\;\;\;\;\;\;\;\;\;\;\;\;\;\;\;\;\;\;  
\times \prod_{l=1}^{n_{0}} \frac{(x^{-} + \frac{1}{x^{-}} - \tilde{\lambda}_{l} + \frac{3i}{2g} )}{(x^{-} + \frac{1}{x^{-}} - \tilde{\lambda}_{l} + \frac{i}{2g} ) }
\frac{(x^{-} + \frac{1}{x^{-}} + \tilde{\lambda}_{l} + \frac{3i}{2g} )}{(x^{-} + \frac{1}{x^{-}} + \tilde{\lambda}_{l} + \frac{i}{2g} ) }  \nonumber \\
&&+ \frac{(x^{+} + x^{-}) [ 1 + (x^{-})^2 ]}{2 x^{+} ( 1 + x^{+} x^{-} )}
\prod_{j=1}^{N} \frac{x^{-}_{j}}{x^{+}_{j}} \frac{(x^{+} x^{-}_{j} - 1)( x^{+} - x^{+}_{j} ) ( x^{-} + x^{+}_{j} ) (x^{+} x^{+}_{j} + 1)}{(x^{-} x^{-}_{j} - 1)( x^{+} - x^{-}_{j} ) ( x^{+} + x^{+}_{j} ) (x^{+} x^{-}_{j} + 1)} \nonumber \\
&& \;\;\;\;\;\;\;\;\;\;\;\;\;\;\;\;\;\;\;\;\;\;\;\;\;\;\;\;\;\;\;\;\;\;\;\;\;\;\;\;\;\;\;\;\;\;\;\;\;\;\;\;\;\;\;\;\;\;\;\;\;\;\;\;\;\;\;\;\;\;\;\;\;\;\;  
\times \prod_{k=1}^{m_{0}} \frac{x^{+}}{x^{-}} \frac{[1 - ( x^{-} z^{-}_{k})^{2}]}{[1 - ( x^{+} z^{-}_{k})^{2}]}
\end{eqnarray}
provided that the set of variables $\{ z^{\pm}_{1} , \dots , z^{\pm}_{m_{0}} \}$ and $\{ \tilde{\lambda}_{1} , \dots , \tilde{\lambda}_{n_{0}} \}$ satisfy
the following system of Bethe ansatz equations \footnote{The author thanks R. Nepomechie for pointing out sign typos in the Eq. (\ref{ba2}) of a previous version.},
\begin{eqnarray}
\label{ba1}
\prod_{j=1}^{N} \frac{(z^{-}_{k} + x^{-}_{j})}{(z^{-}_{k} - x^{-}_{j})} \frac{(z^{-}_{k} - x^{+}_{j})}{(z^{-}_{k} + x^{+}_{j})} \Theta(z^{\pm}_{k}) &=&
\prod_{j=1}^{n_{0}}
\frac{(z^{-}_{k} + \frac{1}{z^{-}_{k}} - \tilde{\lambda}_{j} - \frac{i}{2g} )}{(z^{-}_{k} + \frac{1}{z^{-}_{k}} - \tilde{\lambda}_{j} + \frac{i}{2g} )} 
\frac{(z^{-}_{k} + \frac{1}{z^{-}_{k}} + \tilde{\lambda}_{j} - \frac{i}{2g} )}{(z^{-}_{k} + \frac{1}{z^{-}_{k}} + \tilde{\lambda}_{j} + \frac{i}{2g} )} \nonumber \\
&& \;\;\;\;\;\;\;\;\;\;\;\;\;\;\;\;\;\;\;\;\;\;\;\;\;\;\;\;\;\;\;\;\;\;\;\;\; k=1, \dots , m_{0} 
\end{eqnarray}
\begin{eqnarray}
\label{ba2}
\prod_{j=1}^{m_{0}} 
\frac{( \tilde{\lambda}_{k} -z^{-}_{j} - \frac{1}{z^{-}_{j}} + \frac{i}{2g} )}{( \tilde{\lambda}_{k} -z^{-}_{j} - \frac{1}{z^{-}_{j}} - \frac{i}{2g} )} 
\frac{( \tilde{\lambda}_{k} +z^{-}_{j} + \frac{1}{z^{-}_{j}} + \frac{i}{2g} )}{( \tilde{\lambda}_{k} +z^{-}_{j} + \frac{1}{z^{-}_{j}} - \frac{i}{2g} )} &=&
\prod_{\stackrel{j=1}{j \neq k}}^{n_{0}}
\frac{( \tilde{\lambda}_{k} - \tilde{\lambda}_{j}  + \frac{i}{g} )}{( \tilde{\lambda}_{k} - \tilde{\lambda}_{j}  - \frac{i}{g} )} 
\frac{( \tilde{\lambda}_{k} + \tilde{\lambda}_{j}  + \frac{i}{g} )}{( \tilde{\lambda}_{k} + \tilde{\lambda}_{j}  - \frac{i}{g} )} \nonumber \\
&& \;\;\;\;\;\;\;\;\;\;\;\;\;\;\;\;\;\;\; k=1, \dots , n_{0} 
\end{eqnarray}
The function $\Theta(z^{\pm})$ is given by \footnote{As observed in \cite{opb}, the function $\Theta(z^{\pm})$ simplifies to unity due to the constraint (\ref{eqv}). }
\begin{equation}
\label{teta}
\Theta(z^{\pm}) =  \frac{2 z^{+} z^{-}(z^{+} + \frac{1}{z^{+}} - \frac{i}{2g} )}{(z^{+} +  z^{-})( z^{+} z^{-} + 1)}
\end{equation}
which contains contribution from the boundaries. 

As discussed in \cite{FEND,NEPO1} the eigenvalues of the double-row transfer matrix are an important constituent of the momenta quantization rule
for particles on the half line with boundaries. In the next section we shall make use of these eigenvalues to derive asymptotic Bethe ansatz equations 
describing the spectrum of open strings attached to maximal giant gravitons.

\section{Asymptotic Bethe ansatz equations}
In the Ref. \cite{HM} the authors generalized the formalism of magnon scattering in the planar limit of the AdS/CFT correspondence
to include the presence of boundaries. In particular, the authors investigated the open boundary conditions 
associated to open strings in $AdS_{5} \times S^{5}$ attached to D$3$-branes also known as maximal giant gravitons. 
In this sense, the scattering of a particle with the boundaries is described by the so called boundary $S$-matrices or 
reflection matrices, and compatibility between the boundary scattering and the bulk integrability demands the reflection
matrices to satisfy the reflection equations (\ref{rem}, \ref{rep}).
From the perspective of the $AdS_{5} \times S^{5}$ string theory, the scattering amplitudes of the world-sheet excitations are
described by a $S$-matrix invariant relative to the centrally extended $su(2|2) \otimes su(2|2)$ superalgebra \cite{tba1,tba1a}. 
The corresponding $S$-matrix is fully constrained by this symmetry algebra, up to an overall multiplicative scalar factor, and it
is explicitly given by
\begin{equation}
\label{ss}
\hat{\mathcal{S}}(\lambda_{1} , \lambda_{2}) = S_{0}(\lambda_{1} ,\lambda_{2}) ^2 \; S(\lambda_{1} ,\lambda_{2}) \otimes S(\lambda_{1} ,\lambda_{2})
\end{equation}
where each term $S(\lambda_{1} ,\lambda_{2})$ is one copy of the centrally extended $su(2|2)$ $S$-matrix given in the section $2$.
The overall scalar factor $S_{0}(\lambda_{1} ,\lambda_{2})$ has been investigated by many authors, see for instance \cite{fase}, but
here we would like to proceed without making any assumption concerning its explicit form.

According to Hofman and Maldacena \cite{HM}, the $Y=0$ giant graviton brane is described by a $su(1|2) \otimes su(1|2)$ theory.
More precisely, the reflection matrices characterizing such boundaries are formed by two copies of the $su(1|2)$ reflection matrix
\begin{equation}
\label{kk}
\mathcal{K}^{\pm}(\lambda) = \left[ k_{0}^{\pm}(\lambda) \right]^2 K^{\pm}(\lambda) \otimes K^{\pm}(\lambda) 
\end{equation}
where each term $K^{\pm}(\lambda)$ consist of the reflection matrix described in the section $3$.
For integrable systems on the half line with open boundaries the reflection equation determines the possible $K$-matrices preserving 
integrability up to an overall multiplicative scalar factor $k_{0}^{\pm}(\lambda)$. If we restrict our discussion only to open spin chains
associated to the reflection matrix, the phase factors $k_{0}^{\pm}(\lambda)$ are not of relevance. However, the
complete determination of the boundary $S$-matrix for the physical excitations of the integrable theory requires the determination of this
scalar factor which contains the data about possible boundary bound states. The computation of the boundary phase factors
$k_{0}^{\pm}(\lambda)$ was recently addressed in the Ref. \cite{tba2} but in what follows we shall not assume any particular form for them.

The purpose of this section is to derive the momenta quantization rule for magnons interacting on a half line through the centrally extended
$su(2|2) \otimes su(2|2)$ factorizable $S$-matrix (\ref{ss}), whose interaction with the boundaries is mediated by the
$su(1|2) \otimes su(1|2)$ boundary $S$-matrix (\ref{kk}). The asymptotic regime of open strings excitations on giant 
gravitons is described  by such bulk and boundary $S$-matrices, which can be viewed as a magnon propagating on an inhomogeneous spin chain
that bounces off a wall and changes its momenta from $\lambda$ to $-\lambda$ \cite{HM}. In order to derive such quantization rule
we shall consider the asymptotic Bethe ansatz framework \cite{tba3} where the conservation of the number of particles as well as their
asymptotic momenta is justified by the existence of a complete set of conserved charges. This quantization rule for the asymptotic momenta
of a system with $N$ particles on an interval of length $L$ has been discussed in \cite{FEND,NEPO1}, and similarly to the case with periodic
boundary conditions \cite{MELO}, the double-row transfer matrix eigenvalues are a fundamental constituent. It turns out that the momenta
$\lambda_{k}$ are constrained by the following relation, 
\begin{equation}
e^{-2 i \lambda_{k} L} = \frac{\hat{\Lambda}(\lambda = \lambda_{k}, \{\lambda_{j} \})}{\psi^{+}(\lambda_{k}) \psi^{-}(\lambda_{k})}
\end{equation}
where $\hat{\Lambda}(\lambda, \{\lambda_{j} \})$ consist of the eigenvalues of the double row transfer matrix built from (\ref{ss}) 
and (\ref{kk}). As pointed out in \cite{FEND}, this quantization rule can be derived with the successive application of the Faddeev-Zamolodchikov algebra enjoyed by the 
$S$-matrix \cite{PLEF} in association with its boundary counterpart \cite{NEPOa}. The functions $\psi^{\pm}(\lambda)$ arise from the identities
\begin{eqnarray}
\mbox{Str}_{1} \left[  \mathcal{R}_{2 1} (\lambda, -\lambda) 
\mathcal{K}^{+}_{1} (\lambda) \right] = \psi^{-}(\lambda) \mathcal{K}_{2}^{+}(-\lambda) \;\; \mbox{and} \;\;
\hat{\mathcal{S}}(\lambda , \lambda) &=&  \psi^{+}(\lambda) P \otimes P \; ,
\end{eqnarray}
and they are given by
\begin{eqnarray}
\psi^{-} (\lambda) &=& \left[ \frac{k^{+}_{0}(\lambda)  S_{0} (\lambda, -\lambda )}{k^{+}_{0}(-\lambda)} \frac{(x^{+} + x^{-})[1 + (x^{+})^2]}{2 x^{+} (1 + x^{+} x^{-})} \right]^2 \nonumber \\
\psi^{+} (\lambda) &=& S_{0} (\lambda, \lambda )^2 \; .
\end{eqnarray}
Here we have also defined the $R$-matrix $\mathcal{R}(\lambda_{1}, \lambda_{2}) = \left( P \otimes P \right) \hat{\mathcal{S}}(\lambda_{1}, \lambda_{2})$.

Due to the tensor product structure of the bulk and boundary $S$-matrices we can read the eigenvalues  
$\hat{\Lambda}(\lambda, \{\lambda_{j} \})$ directly from the spectrum derived in the section $4$. For instance they are given by
\begin{equation}
\label{asyeig}
\hat{\Lambda}(\lambda, \{\lambda_{j} \}) = \left[ k_{0}^{+}(\lambda) k_{0}^{-}(\lambda) \right]^2
\prod_{j=1}^{N} \left[ S_{0}(\lambda,\lambda_{j}) S_{0}(\lambda_{j}, -\lambda) \right]^2
\Lambda(\lambda, \{ \lambda_{j} \} )
\Lambda^{\prime} (\lambda, \{ \lambda_{j} \} )
\end{equation}
where $\Lambda(\lambda, \{ \lambda_{j} \} )$ consist of the eigenvalues (\ref{eigl}) parameterized by Bethe roots
$\{ z_{1,j}^{\pm} , \tilde{\lambda}_{1,j} \}$ and $\Lambda^{\prime} (\lambda, \{ \lambda_{j} \} )$ corresponds to the 
eigenvalues associated to the second $su(2|2)$ copy given in terms of Bethe roots $\{ z_{2,j}^{\pm} , \tilde{\lambda}_{2,j} \}$.
Now considering the explicit expressions (\ref{eigl}), (\ref{ba1}), (\ref{ba2}) and (\ref{asyeig}), one finds the following set
of nested Bethe ansatz equations for the asymptotic magnon momenta,
\begin{eqnarray}
\label{finalba}
\left[ \frac{x^{+}_{k}}{x^{-}_{k}}  \right]^{-2(L + N - \frac{m_{1}}{2} - \frac{m_{2}}{2} )} \Phi(\lambda_{k}) &=& 
\prod_{\stackrel{j=1}{j \neq k}}^{N} \left[ S_{0}(\lambda_{k}, \lambda_{j}) S_{0}(\lambda_{j}, -\lambda_{k}) \frac{(x^{-}_{k} + x^{-}_{j})(x^{-}_{k} - x^{+}_{j})}{(x^{+}_{k} - x^{-}_{j})(x^{+}_{k} + x^{+}_{j})} \right]^2  \nonumber \\
&\times & \prod_{\alpha=1}^{2} \prod_{l=1}^{m_{\alpha}}  \frac{ (x^{+}_{k} -  z^{-}_{\alpha,l} ) (x^{+}_{k} +  z^{-}_{\alpha,l} )}{( x^{-}_{k}  - z^{-}_{\alpha,l} )( x^{-}_{k}  + z^{-}_{\alpha,l} ) }
\end{eqnarray}
\begin{eqnarray}
\label{finalba1}
\prod_{j=1}^{N} \frac{(z^{-}_{\alpha, k} + x^{-}_{j})}{(z^{-}_{\alpha, k} - x^{-}_{j})} \frac{(z^{-}_{\alpha , k} - x^{+}_{j})}{(z^{-}_{\alpha , k} + x^{+}_{j})} \Theta(z^{\pm}_{\alpha , k}) &=&
\prod_{j=1}^{n_{\alpha}}
\frac{(z^{-}_{\alpha , k} + \frac{1}{z^{-}_{\alpha , k}} - \tilde{\lambda}_{\alpha , j} - \frac{i}{2g} )}{(z^{-}_{\alpha , k} + \frac{1}{z^{-}_{\alpha , k}} - \tilde{\lambda}_{\alpha , j} + \frac{i}{2g} )} 
\frac{(z^{-}_{\alpha , k} + \frac{1}{z^{-}_{\alpha , k}} + \tilde{\lambda}_{\alpha , j} - \frac{i}{2g} )}{(z^{-}_{\alpha , k} + \frac{1}{z^{-}_{\alpha , k}} + \tilde{\lambda}_{\alpha , j} + \frac{i}{2g} )} \nonumber \\
&& \;\;\;\;\;\;\;\;\;\;\;\;\;\;\;\;\;\;\;\;\;\;\;\;\;\;\;\;\;\;\;\;\;\;\;\;\; \alpha =1,2 \;\;\;\;\;\;\; k=1, \dots , m_{\alpha} \nonumber \\
\end{eqnarray}
\begin{eqnarray}
\label{finalba2}
\prod_{j=1}^{m_{\alpha}} 
\frac{( \tilde{\lambda}_{\alpha , k} -z^{-}_{\alpha ,j} - \frac{1}{z^{-}_{\alpha ,j}} + \frac{i}{2g} )}{( \tilde{\lambda}_{\alpha ,k} -z^{-}_{\alpha ,j} - \frac{1}{z^{-}_{\alpha ,j}} - \frac{i}{2g} )} 
\frac{( \tilde{\lambda}_{\alpha ,k} +z^{-}_{\alpha ,j} + \frac{1}{z^{-}_{\alpha ,j}} + \frac{i}{2g} )}{( \tilde{\lambda}_{\alpha ,k} +z^{-}_{\alpha ,j} + \frac{1}{z^{-}_{\alpha ,j}} - \frac{i}{2g} )} &=&
\prod_{\stackrel{j=1}{j \neq k}}^{n_{\alpha}}
\frac{( \tilde{\lambda}_{\alpha ,k} - \tilde{\lambda}_{\alpha ,j}  + \frac{i}{g} )}{( \tilde{\lambda}_{\alpha ,k} - \tilde{\lambda}_{\alpha ,j}  - \frac{i}{g} )} 
\frac{( \tilde{\lambda}_{\alpha ,k} + \tilde{\lambda}_{\alpha ,j}  + \frac{i}{g} )}{( \tilde{\lambda}_{\alpha ,k} + \tilde{\lambda}_{\alpha ,j}  - \frac{i}{g} )} \nonumber \\
&& \;\;\;\;\;\;\;\;\;\;\; \alpha =1,2 \;\;\;\;\;\;\; k=1, \dots , n_{\alpha} \nonumber \\ 
\end{eqnarray}
The function $\Theta(z^{\pm})$ is determined by the Eq. (\ref{teta}) while $\Phi(\lambda)$ is given by
\begin{equation}
\Phi(\lambda) = \left[ \left( \frac{x^{+}}{x^{-}} \right)^2 \frac{1}{k^{+}_{0}(-\lambda) k^{-}_0 (\lambda)}  \right]^2
\end{equation} 
which contains dependence on the boundary phase factors. As it occurs in the periodic case \cite{MELO}, the Eq. (\ref{finalba}) suggests that 
$L^{\prime} = L + N - \frac{m_{1}}{2} - \frac{m_{2}}{2} $ should play the role of effective scale encoding the 
angular momenta of the $AdS_5 \times S^5$ theory in the light-cone gauge \cite{ARFR}. Concerning the dependence of the function $\Phi(\lambda)$
with the boundary phase factors $k^{\pm}_{0}(\lambda)$, we close this section remarking that according to the crossing relation
analysis of \cite{tba2} and the strong coupling study performed in \cite{HM}, the boundary phase factors depend explicitly on
the bulk phase factor $S_{0} (\lambda, -\lambda)$, which is in qualitative agreement with the analysis performed in \cite{LOB} for long 
range open spin chains.

\section{Spin chain hamiltonian}
Nowadays it is well known that integrable spin chains with open boundaries can be 
obtained in the first order expansion of Sklyanin's double-row transfer matrix \cite{SK}.
However, the standard derivation of such spin chains
considers homogeneous transfer matrices whose bulk $S$-matrix and reflection matrices exhibit certain
properties.
In order to derive such spin chain hamiltonians it is usually evoked the regularity of the formers,
\begin{eqnarray}
S_{12}(\lambda_{0} , \lambda_{0}) &\sim & P_{12} \nonumber \\
K^{-} (\lambda_{0}) &\sim & I ,
\end{eqnarray}
at a certain value $\lambda_{0}$ of the spectral parameter. We recall here that $P$ stands for the permutation operator while
$I$ denotes the identity matrix.

The reflection matrix considered here does not exhibit such property and this fact makes necessary to generalize the
mapping proposed by Sklyanin in \cite{SK}. Moreover, in the case considered here the bulk $S$-matrix does not depend only on
the difference of the spectral parameters and this feature leads us to choose appropriate values of
$\lambda_{0}$ in order to obtain suitable spin chain hamiltonians.
Here we find that a spin chain hamiltonian with open boundaries can be obtained from non-regular reflection matrices by considering the logarithmic derivative 
of the transfer matrix at a certain point $\lambda_{0}$. Here we shall consider the point $\lambda_{0}=\pi$ where the $S$-matrix satisfy the following
property
\begin{equation}
\label{ppi}
S_{12}(\pi,\pi) = S_{21}(\pi, -\pi) = - P_{12}.
\end{equation}
The property (\ref{ppi}) allows us to show that the double-row transfer matrix (\ref{tm}) with
inhomogeneities $\lambda_{j}=\lambda_{0}=\pi$ is given by
\begin{equation}
\label{t0}
T(\pi, \{ \pi \} ) = \mbox{Str}_{\mathcal{A}} \left[K_{\mathcal{A}}^{+} (\pi) \right] K_{1}^{-} (\pi),
\end{equation}
and that the derivative of the transfer matrix at the point $\lambda = \pi$ can be written as
\begin{eqnarray}
\label{dt0}
&& \left. \frac{dT(\lambda, \{ \pi \} )}{d\lambda} \right|_{\lambda = \pi} =
- \sum_{j=2}^{N-1} \mbox{Str}_{\mathcal{A}} \left[ K_{\mathcal{A}}^{+}(\pi) \right] h_{j,j+1} K_{1}^{-}(\pi) 
- \sum_{j=2}^{N-1} \mbox{Str}_{\mathcal{A}} \left[ K_{\mathcal{A}}^{+}(\pi) \right] \bar{h}_{j,j+1} K_{1}^{-}(\pi) \nonumber \\ 
&&+\mbox{Str}_{\mathcal{A}} \left[ K_{\mathcal{A}}^{\prime +}(\pi) \right] K_{1}^{-}(\pi) 
- \mbox{Str}_{\mathcal{A}} \left[ K_{\mathcal{A}}^{+}(\pi) h_{N, \mathcal{A}} \right] K_{1}^{-}(\pi) 
- \mbox{Str}_{\mathcal{A}} \left[ K_{\mathcal{A}}^{+}(\pi) \right] h_{1,2} K_{1}^{-}(\pi) \nonumber \\
&&+ \mbox{Str}_{\mathcal{A}} \left[ K_{\mathcal{A}}^{+}(\pi) \right] K_{1}^{\prime -}(\pi)
- \mbox{Str}_{\mathcal{A}} \left[ K_{\mathcal{A}}^{+}(\pi) \right] K_{1}^{-}(\pi) \bar{h}_{1,2} 
- \mbox{Str}_{\mathcal{A}} \left[ K_{\mathcal{A}}^{+}(\pi) \bar{h}_{N, \mathcal{A}} \right] K_{1}^{-}(\pi). \nonumber \\
\end{eqnarray}
In the expression (\ref{dt0}) the terms $K^{\prime \pm}(\pi)$ denote the first derivative of the reflection matrices
at the point $\lambda = \pi$, i.e. $K^{\prime \pm}(\pi) = \left. \frac{d K^{\pm} (\lambda)}{d\lambda} \right|_{\lambda = \pi}$,
while the two site hamiltonians $h_{i,j}$ and $\bar{h}_{i,j}$ are defined as
\begin{eqnarray}
h_{i,j} = P_{ij} \left. \frac{dS_{ij}(\lambda, \pi)}{d\lambda} \right|_{\lambda = \pi} \;\;\;\;\;\; \mbox{and} \;\;\;\;\;\;
\bar{h}_{i,j} = P_{ij} \left. \frac{dS_{ij}(\pi, -\lambda)}{d\lambda} \right|_{\lambda = \pi} .
\end{eqnarray}
It is also important to emphasize here that we have only used the permutator algebra in order to obtain the relations (\ref{t0}) and
(\ref{dt0}).

With the above considerations the spin chain hamiltonian defined as $\mathcal{H} = \left. \frac{d \ln T(\lambda, \{ \pi \} )}{d\lambda} \right|_{\lambda = \pi} = 
T(\pi, \{ \pi \} )^{-1} \left. \frac{dT(\lambda, \{ \pi \} )}{d\lambda} \right|_{\lambda = \pi}$ turns out to be given by
\begin{eqnarray}
\label{ham}
\mathcal{H} &=& - \sum_{j=1}^{N-1} h_{j,j+1} - \sum_{j=1}^{N-1} \bar{h}_{j,j+1}  - \left( K^{-}_{1}(\pi) \right)^{-1} \left[ h_{1,2} , K^{-}_{1}(\pi) \right]
+ \left( K^{-}_{1}(\pi) \right)^{-1} K_{1}^{\prime -}(\pi) \nonumber \\
&-& \frac{\mbox{Str}_{\mathcal{A}} \left[ K_{\mathcal{A}}^{+}(\pi) h_{N, \mathcal{A}} \right]}{\mbox{Str}_{\mathcal{A}} \left[ K_{\mathcal{A}}^{+}(\pi) \right]} - \frac{\mbox{Str}_{\mathcal{A}} \left[ K_{\mathcal{A}}^{+}(\pi) \bar{h}_{N, \mathcal{A}} \right]}{\mbox{Str}_{\mathcal{A}} \left[ K_{\mathcal{A}}^{+}(\pi) \right]} + \frac{\mbox{Str}_{\mathcal{A}} \left[ K_{\mathcal{A}}^{\prime +}(\pi) \right]}{\mbox{Str}_{\mathcal{A}} \left[ K_{\mathcal{A}}^{+}(\pi) \right]}.
\end{eqnarray}

At this point it is worthwhile to make some comments concerning the hamiltonian $\mathcal{H}$. In the case considered here the bulk terms satisfy
$h_{i,j} = \bar{h}_{i,j}$ but the derivation of (\ref{ham}) does not rely on this property. In contrast to the standard hamiltonians derived from
regular solutions of the reflection equation, the main feature of the hamiltonian (\ref{ham}) is the residual interaction between the bulk
term $h_{1,2}$ and the boundary element $K_{1}^{-}(\pi)$. We remark here that integrable spin chains with interactions between the bulk term $h_{1,2}$ and boundary terms
acting in the site $1$ had appeared previously in the context of gauge theories describing the one-loop anomalous dimensions of giant 
gravitons in the $SO(6)$ sector of the $\mathcal{N}=4$ Super Yang-Mills \cite{intro12}. Although the terms appearing in \cite{intro12} do not seem
to be equivalent to the ones in (\ref{ham}), it would be interesting to investigate if such kind of spin chain with open boundaries can arise
from non-regular solutions of the reflection equations or more general realizations of the reflection algebras. With respect to the cases where
the reflection matrix is regular, it is straightforward to see that the hamiltonian (\ref{ham}) reduces to the standard form originally derived in \cite{SK}.
 
\section{Concluding Remarks}
This work is mainly concerned with the derivation of the Bethe ansatz equations for particles interacting through a
$S$-matrix invariant with respect to the centrally extended $su(2|2)$ Lie superalgebra, whose interactions with the
boundaries belong to a smaller symmetry algebra, namely the superalgebra $su(1|2)$.

The $su(1|2)$ reflection matrices are derived from a quantum group like approach introduced in \cite{DELMAC}, which provides
a more solid ground for the method used by Hofman and Maldacena \cite{HM}. The associated double-row
inhomogeneous transfer matrix was diagonalized by means of the algebraic Bethe ansatz which allowed us to derive the 
quantization rule on the half line for the asymptotic momenta of magnons interacting through the 
$su(2|2) \otimes su(2|2)$ $S$-matrix and reflected by a rigid wall described by a  $su(1|2) \otimes su(1|2)$ boundary
$S$-matrix. These asymptotic Bethe ansatz equations may be of importance for the study of the spectrum of open strings in 
$AdS_{5} \times S^5$ background attached to the called $Y=0$ giant graviton brane in the
thermodynamic limit.

Here we have considered only the reflection matrix associated to the $Y=0$ giant graviton brane. However, the $Z=0$ case described by
a $su(2|2) \otimes su(2|2)$ reflection matrix was also studied in the Ref. \cite{HM} and it would be interesting to investigate if it
can be approached in the same fashion, as well as their $q$-deformed cases presented in \cite{qdef}.

Although the centrally extended $su(2|2)$ $S$-matrix is equivalent to Shastry's $R$-matrix, the integrable boundaries considered
here are different from the ones previously obtained for the one-dimensional Hubbard model \cite{ZHOU,WAD}. From the perspective of
the centrally extended $su(2|2)$ Lie superalgebra, these differences could be understood as follows. The energy $\epsilon$ and the momenta
$\lambda$ are elements of the algebra and they are given in terms of the variables $x^{\pm}$ by
\begin{equation}
\epsilon = \frac{i g}{2} \left( x^{-} - \frac{1}{x^{-}} - x^{+} + \frac{1}{x^{+}} \right) \;\;\;\;\; , \;\;\;\; e^{i\lambda} = \frac{x^{+}}{x^{-}}.
\end{equation}
The reflection mapping considered here, $x^{\pm} \rightarrow - x^{\mp}$, corresponds to $(\epsilon, \lambda) \rightarrow (\epsilon, -\lambda)$ while
the reflection mapping adopted in \cite{ZHOU,WAD} is $x^{\pm} \rightarrow - \frac{1}{x^{\mp}}$ due to the
specific parameterization employed. The latter corresponds to $(\epsilon, \lambda) \rightarrow (- \epsilon, \lambda)$ which could be thought
as a particle being converted in an anti-particle when it hits the boundaries. Although the momenta $\lambda$ is not inverted in that case, 
we would like to remark that reflection matrices describing the reflection of a particle with the boundary which comes back as an
anti-particle have been considered previously in the literature \cite{DOI}.

From the point of view of the Quantum Inverse Scattering Method with open boundaries, the reflection matrix
considered here exhibits the peculiar feature of being non-regular. As far as we know, the derivation of
spin chain hamiltonians with open boundaries associated to such reflection matrices was not known in the
literature. For instance, a variety of regular solutions of the reflection equation is known for
$q$-deformed Lie algebras and superalgebras \cite{Regl} and super-Yangians \cite{Regsy}, and we
hope the possibility of deriving integrable open spin chains from non-regular $K$-matrices
presented here to motivate the search for such solutions.

\section{Acknowledgements}
The author thanks A. Agarwal for useful discussions and N. Beisert for valuable suggestions, discussions and for commenting on this
manuscript. The author also thanks R. Nepomechie for discussions and correspondence.

\newpage
\section*{\bf Appendix A: Spectral equivalences}
\setcounter{equation}{0}
\renewcommand{\theequation}{A.\arabic{equation}}
In this appendix we demonstrate the equivalence of our results for the reflection matrices (\ref{km},\ref{kp})
with results previously presented in the literature \cite{HM,NEPO}. We shall make use of 
gauge transformations in order to show that the associated 
double-row transfer matrices are related by a similarity transformation and thus they possess the
same spectrum. 

In order to estabilish this spectral equivalence we firstly recall the definition (\ref{tm}) for the 
double row-transfer matrix,
\begin{equation}
\label{tma}
T(\lambda , \{\lambda_{j} \}) = \mbox{Str}_{\mathcal{A}} \left[ K^{+}_{\mathcal{A}}(\lambda) 
S_{\mathcal{A} N}(\lambda, \lambda_{N}) \dots S_{\mathcal{A} 1}(\lambda, \lambda_{1})
K^{-}_{\mathcal{A}}(\lambda) 
S_{1 \mathcal{A} }(\lambda_{1}, -\lambda) \dots S_{N \mathcal{A} }(\lambda_{N}, -\lambda)  \right].
\end{equation}
Next we proceed by inserting terms $\mathcal{E}_{\mathcal{A}} \mathcal{E}_{\mathcal{A}}^{-1}$ and $\mathcal{F}_{\mathcal{A}} \mathcal{F}_{\mathcal{A}}^{-1}$
in between the elements of the double-row operator (\ref{tma}) in the following way,
\begin{eqnarray}
\label{gaugetm}
T(\lambda , \{\lambda_{j} \}) = && \mbox{Str}_{\mathcal{A}} \left[  K^{+}_{\mathcal{A}}(\lambda) \mathcal{E}_{\mathcal{A}} \mathcal{E}_{\mathcal{A}}^{-1} S_{\mathcal{A} N}(\lambda, \lambda_{N}) \mathcal{E}_{\mathcal{A}} \mathcal{E}_{\mathcal{A}}^{-1} \dots \mathcal{E}_{\mathcal{A}} \mathcal{E}_{\mathcal{A}}^{-1} S_{\mathcal{A} 1}(\lambda, \lambda_{1}) \mathcal{E}_{\mathcal{A}} \mathcal{E}_{\mathcal{A}}^{-1} \right. \nonumber \\
&& \left. K^{-}_{\mathcal{A}}(\lambda) \mathcal{F}_{\mathcal{A}} \mathcal{F}_{\mathcal{A}}^{-1} S_{1 \mathcal{A} }(\lambda_{1}, -\lambda) \mathcal{F}_{\mathcal{A}} \mathcal{F}_{\mathcal{A}}^{-1} \dots \mathcal{F}_{\mathcal{A}} \mathcal{F}_{\mathcal{A}}^{-1} S_{N \mathcal{A} }(\lambda_{N}, -\lambda) \mathcal{F}_{\mathcal{A}} \mathcal{F}_{\mathcal{A}}^{-1} \right] \nonumber \\
\end{eqnarray}
which can be more conveniently rearrenged as
\begin{eqnarray}
\label{gaugetm1}
T(\lambda , \{\lambda_{j} \}) = && \mbox{Str}_{\mathcal{A}} \left[ \left( \mathcal{F}_{\mathcal{A}}^{-1} K^{+}_{\mathcal{A}}(\lambda) \mathcal{E}_{\mathcal{A}} \right)
\left( \mathcal{E}_{\mathcal{A}}^{-1} S_{\mathcal{A} N}(\lambda, \lambda_{N}) \mathcal{E}_{\mathcal{A}} \right) \dots
\left( \mathcal{E}_{\mathcal{A}}^{-1} S_{\mathcal{A} 1}(\lambda, \lambda_{1}) \mathcal{E}_{\mathcal{A}} \right) \right. \nonumber \\
&& \left. \left( \mathcal{E}_{\mathcal{A}}^{-1} K^{+}_{\mathcal{A}}(\lambda) \mathcal{F}_{\mathcal{A}} \right) \left( \mathcal{F}_{\mathcal{A}}^{-1} S_{1 \mathcal{A} }(\lambda_{1}, -\lambda) \mathcal{F}_{\mathcal{A}} \right) \dots \left( \mathcal{F}_{\mathcal{A}}^{-1} S_{N \mathcal{A} }(\lambda_{N}, -\lambda) \mathcal{F}_{\mathcal{A}} \right) \right]. \nonumber \\
\end{eqnarray}

Considering a similarity transformation in the quantum space generated by
\begin{equation}
\mathcal{U} = \bigotimes_{j=1}^{N} \mathcal{E}_{j}
\end{equation}
under the assumption that $\mathcal{E}_{j} = \mathcal{E}_{j}(\lambda_{j})$, $\mathcal{E}_{\mathcal{A}} = \mathcal{E}_{\mathcal{A}}(\lambda)$ and
$\mathcal{F}_{\mathcal{A}} = \mathcal{E}_{\mathcal{A}}(-\lambda)$, we are left with a transformed transfer matrix
$\tilde{T}(\lambda , \{\lambda_{j} \})  = \mathcal{U}^{-1} T(\lambda , \{\lambda_{j} \}) \mathcal{U}$ given by
\begin{equation}
\label{tmagauged}
\tilde{T}(\lambda , \{\lambda_{j} \}) = \mbox{Str}_{\mathcal{A}} \left[ \tilde{K}^{+}_{\mathcal{A}}(\lambda) 
\tilde{S}_{\mathcal{A} N}(\lambda, \lambda_{N}) \dots \tilde{S}_{\mathcal{A} 1}(\lambda, \lambda_{1})
\tilde{K}^{-}_{\mathcal{A}}(\lambda) 
\tilde{S}_{1 \mathcal{A} }(\lambda_{1}, -\lambda) \dots \tilde{S}_{N \mathcal{A} }(\lambda_{N}, -\lambda)  \right],
\end{equation}
where we have defined the elements
\begin{eqnarray}
\label{gg}
\tilde{S}_{\mathcal{A} j}(\lambda, \lambda_{j}) &=& \mathcal{E}_{\mathcal{A}}^{-1} (\lambda) \mathcal{E}_{j}^{-1} (\lambda_{j}) S_{\mathcal{A} j}(\lambda, \lambda_{j}) \mathcal{E}_{\mathcal{A}}(\lambda) \mathcal{E}_{j}(\lambda_{j}) \nonumber \\
\tilde{K}^{+}_{\mathcal{A}}(\lambda) &=& \mathcal{E}_{\mathcal{A}}^{-1} (-\lambda) K^{+}_{\mathcal{A}}(\lambda) \mathcal{E}_{\mathcal{A}}(\lambda) \\
\tilde{K}^{-}_{\mathcal{A}}(\lambda) &=& \mathcal{E}_{\mathcal{A}}^{-1} (\lambda) K^{-}_{\mathcal{A}}(\lambda) \mathcal{E}_{\mathcal{A}}(-\lambda). \nonumber
\end{eqnarray}
One can verifies that the set of transformations described by the relations (\ref{gg}) preserves both the Yang-Baxter relation (\ref{YB})
and the reflection equations (\ref{rem}) and (\ref{rep}). Furthermore, by choosing 
\begin{equation}
\mathcal{E}_{\mathcal{A}}(\lambda) = \pmatrix{
e^{i\frac{\lambda}{4}} & 0 & 0 & 0 \cr
0 & 1 & 0 & 0 \cr
0 & 0 & 1 & 0 \cr
0 & 0 & 0 & e^{i\frac{\lambda}{4}} \cr}
\end{equation}
one finds 
\begin{equation}
\tilde{K}^{-}_{\mathcal{A}}(\lambda) = \pmatrix{
-e^{-i\lambda} & 0 & 0 & 0 \cr
0 & 1 & 0 & 0 \cr
0 & 0 & 1 & 0 \cr
0 & 0 & 0 & 1 \cr}
\end{equation}
and $\tilde{K}^{+}_{\mathcal{A}}(\lambda) = \tilde{K}^{-}_{\mathcal{A}}(-\lambda)$, which are precisely the reflection matrices described by Hofman and Maldacena in \cite{HM}.

On the other hand, by setting 
\begin{equation}
\mathcal{E}_{\mathcal{A}}(\lambda) = \pmatrix{
0 & e^{-i\frac{\lambda}{4}} & 0 & 0 \cr
0 & 0 & 0 & 1 \cr
0 & 0 & 1 & 0 \cr
e^{i \frac{3}{4} \lambda} & 0 & 0 & 0 \cr}
\end{equation}
we are left with the transformed reflection matrices 
\begin{equation}
\tilde{K}^{-}_{\mathcal{A}}(\lambda) = \pmatrix{
e^{-i\lambda} & 0 & 0 & 0 \cr
0 & -1 & 0 & 0 \cr
0 & 0 & 1 & 0 \cr
0 & 0 & 0 & 1 \cr}
\end{equation}
and $\tilde{K}^{+}_{\mathcal{A}}(\lambda) = \tilde{K}^{-}_{\mathcal{A}}(-\lambda)$ which coincides with the results of \cite{NEPO}.

\end{document}